\documentstyle[12pt]{article}
\newcommand{\be}{\begin{eqnarray}}
\newcommand{\ee}{\end{eqnarray}}
\newcommand{\nn}{\nonumber}

\newcommand{\pprime}{{\prime\prime}}
\newcommand{\ppprime}{{\prime\prime\prime}}
\newcommand{\V}{{{\cal V}}}
\newcommand{\con}{\mbox{$\,$\rule{1ex}{1pt}\rule{1pt}{1ex}$\,$}}
\newcommand{\dwedge}{\dot{\wedge}}
\newcommand{\brk}[1]{\left[ #1\right]}
\newcommand{\kla}[1]{\left( #1\right)}
\newcommand{\til}{\tilde{~}}
\newcommand{\x}{\scriptstyle}
\begin{document}
%
%
%
\title{Vertex Normalordering as a Consequence of Nonsymmetric
Bilinearforms in Clifford Algebras}
\author{Bertfried~Fauser\\
Institut f\"ur Theoretische Physik \\
Universit\"at T\"ubingen\\
Auf der Morgenstelle 14
\thanks{e-mail: Bertfried.Fauser@uni-tuebingen.de}}
\date{April, 1995}
\maketitle
%
%
\begin{abstract}
\begin{sloppypar}
arch-ive/9504055

\noindent
We consider Clifford algebras with nonsymmetric bilinear forms,
which are isomorphic to the standard symmetric ones, but not
equal. Observing, that the content of physical theories is
dependent on the injection $\oplus^n\bigwedge \V^{(n)}\rightarrow
CL({\cal V},Q)$ one has to transform to the standard
construction. The injection is of course dependent on the
antisymmetric part of the bilinear form. This process results in
the appropriate vertex normalordering terms, which are now
obtained from the theory itself and not added ad hoc via a
regularization argument.
\end{sloppypar}
\end{abstract}
\section{Introduction}

Nonlinear spinor equations play an important role in several
branches of physics. They appear in high energy, nuclear or
solid state physics. The most recent examples are the
Heisenberg, Nambu, Jona-Lasinio like models \cite{Hei,Nam} of
elementary particle theory or nuclear physics. Even nonlinear
sigma models bear an analogous structure \cite{sigma}. In solid state
physics the Hubbard model \cite{Frad} is a widespread theoretical
tool in describing phenomena from super conductivity up to spin
chains and so on.

The general structure of such models is of the form kinetic term
equals a cubic interaction term with an arbitrary Vertex.
\be
(\sum i\gamma^{\mu}\partial_{\mu}-m)_{II^\prime}\Psi_{I^\prime}&=&
gV_{I I^\prime I^{\pprime} I^{\ppprime}}\Psi_{I^\prime}\Psi_{I^\pprime}
\Psi_{I^\ppprime}.
\ee
Here $\sum\gamma^\mu\partial_\mu$ is the Dirac--operator,
with euclidean or lorentzian signature. The mass could
be zero. With the multi index $I=\{K,\Lambda\}$ we represent the
spinor and its adjoint by $\Lambda$ and the other algebraic and
spatio temporal indices by $K$. If we have fixed an adjoint
spinor, the quadratic form of the Clifford algebra is already determined.
A suitable quantization procedure also has to be applied.

There  are several problems with these equations, which we want
to consider now.
\begin{itemize}
\item[i)] The equations are not renormalizable, because g will
be in general a dimensional quantity.

\item[ii)] In order to define a unique quantization procedure
one has to introduce an ordering, as say timeordering or on a
space like hypersurface the (anti)symmetric one.

\item[iii)] In the language of diagrams, you have to consider
only connected ones, by introducing a normalordering procedure.
\end{itemize}

For point one, there seems to be no principal problem in solid
state physics, because there may be a physically motivated cut
off at the Brillouin zone. In the case of particle physics,
there are several ad hoc regularizations. An approach to these
topics will be given elsewhere \cite{Diss}.

The second point is usually solved by using causality arguments,
which introduce a natural ordering in the polynomials of
the fields at hand. Therefore timeordered products are used in
the covariant formulation, see for example \cite{Itz}. If we
would prefer the Hamilton formalism, the (anti)symmetric
ordering could be used.

At this stage the definition process of the theory should stop,
because including a somehow given Vacuum state\footnote{
One should be able to calculate this vacuum state in a nonlinear
theory, by solving the dynamical problem at hand in a sort of
self consistent problem.} $\vert 0>_{Ph}$, which gives the
representation of the field operators $\prod_{Ph}(\Psi_{I})$, all
quantities are formally\footnote{Formally, because one has to show
uniqueness and existence of the defined objects too, which is a
nontrivial problem.} defined.

Now the third point causes trouble, because the transformation to
the connected components yields infinities. In Fock space\footnote{
The Fock space is only appropriate in free theories or in
perturbation theory, which proves not useful in our case. A
nonperturbative treatment is given in \cite{StBo}.}
this transformation equals a Wick--Dyson
normalordering of the vertex \cite{Lee}. This process results in
the desired connected amplitudes and so called contractions,
which for bilinear and higher order terms yields singularities
at least on the lightcone. In quantum theory these contractions
are related to the finite ground state energies, which when field
quantized become infinite. In this way, and by the convention
that the vacuum has no nonzero quantum numbers a vertex
regularization is also introduced. The field equations read now
\be
(\sum i\gamma^{\mu}\partial_{\mu}-m)_{II^\prime}\Psi_{I^\prime}&=&
gV_{I I^\prime I^{\pprime} I^{\ppprime}}
{\bf :}\Psi_{I^\prime}\Psi_{I^\pprime}\Psi_{I^\ppprime}{\bf :}.
\ee
With the physical propagator
\be
P_{I I^\prime}&:=&{}_{Ph}<0\vert {\cal T}
(\Psi_{I}\Psi_{I^\prime})\vert 0>_{Ph}
\ee
the vertex term changes to
\be
{\bf :}\Psi_{I^\prime}\Psi_{I^\pprime}\Psi_{I^\ppprime}{\bf :}&=&
       \Psi_{I^\prime}\Psi_{I^\pprime}\Psi_{I^\ppprime}
       +P_{I^\prime I^\pprime}\Psi_{I^\ppprime}
       -P_{I^\prime I^\ppprime}\Psi_{I^\pprime}
       +P_{I^\pprime I^\ppprime}\Psi_{I^\prime}.
\ee
But this procedure is nothing but a shift of the problem from
one equation into the other. With this definition the timeordered
equation becomes singular, and hence the whole theory is ill defined.

In this note we want to show, how an embedding of the theory in a
Clifford algebra structure can overcome this problem. Therefore
we consider nonsymmetric bilinearforms and the associated
Clifford algebras. The transformation from such algebras to the
symmetric ones is an isomorphism, but the linear space of
antisymmetric p-vectors is moved. As they carry the physical
information, this is therefore altered.

\section{Clifford Algebras with nonsymmetric bilinearforms}

The Clifford Algebra entered physics with Pauli and Dirac
\cite{Dir}, who used it to linearize the D'Alambertian. So we
should learn more about this procedure. Let $Q$ be a nondegenerate
quadratic form, $\V$ a vector space over $R$ or $C$, then the
Clifford map is an injection from $\V$ into $CL(\V,Q)$ with the
property that every square of a vector element of the Clifford
Algebra is a scalar.
\be
\gamma\quad :\quad V &\rightarrow& CL(\V,Q),\quad e_i\mapsto\gamma_i \nn \\
&&x^2=x \cdot x =Q(x) \in (R,C)
\ee
With linearization we have on a generating set of $\V$
\be
(e_i+e_j)(e_i+e_j)&=&e_i^2+e_j^2+e_i e_j+e_j e_i\nn \\
e_i e_j+e_j e_i&=&Q(e_i+e_j)-Q(e_i)-Q(e_j)=:2G(e_i,e_j)\quad \in
(R,C).
\ee
It is evident from this calculation, that the bilinear form $G$
is symmetric. The whole algebra is now generated from reduced
products of one-vectors. Let $N$ be the set of ordered
partitions of n pieces, $\vert\alpha\vert$ the cardinality of
such a subset, and include the empty set. We define
$1_A=e_{0}$, then an algebra element read
\be
A&:=&\sum_{\alpha\in N} a_{\alpha} e_{\alpha}=A_0+A_1\ldots +A_n \nn \\
e_{\alpha}&:=&e_{i_1}\wedge e_{i_2}\wedge
\ldots\wedge e_{i_r},\quad i_1<i_2 \ldots <i_r,\quad
\vert\alpha\vert=r\,\in N.
\ee
The wedge product means antisymmetric multiplication as in the
Grassmann case. Indeed as a linear space these two constructions
are identical. Thereby the Clifford algebra has the direct
sum decompositions
\be
CL(V,Q)&=&CL_+\oplus CL_-\quad \mbox{as algebra, and}\nn \\
CL(V,Q)&=&\oplus^n \bigwedge \V^{(n)}\quad \mbox{as linear space.}
\ee
But the product intermingles the grades. Let $<\quad>_r$ be the
projector to the homogeneous part of grade $r$, then one has
\be
A_rB_s&=&<AB>_{\vert r-s\vert}+<AB>_{\vert r-s+2\vert}+\ldots +<AB>_{r+s},
\ee
were in the Grassmann case $A_rB_s=AB_{r+s}$ results.

Physicists consider the anticommuting elements of grade $r$ as eg.
scalars, spinors (vectors) , spintensors, (tensors) and so on.
That is, the physical content of the theory is coded explicitly
in this structure.

Now let us see, in which way it is possible to introduce
nonsymmetric bilinear forms. It is obvious, that we have to
leave the above construction, in favor of a more general one.
This can be done by introducing algebra derivatives as proposed
by Chevaley and Riesz \cite{Chev,Riesz}.

First of all, we introduce two more algebraic constructions for
further use. An involution\footnote{
This property is sometimes called conjugation.}
$J$ of period two and the Reversion
$\tilde{~}$ by the rules
\be
J &:& J^2:=id_A\nn \\
   && J(XY):=J(X)J(Y)\nn \\
   && J(R,C):=(R,\bar{C})\nn \\
\tilde{~} &:& <\tilde{~}>_{0+1}:=id_{A_0+A_1}\nn \\
   && (XY)\tilde{~}:=\tilde{Y}\tilde{X}.
\ee
Now we may introduce the desired formulae
\be
a\con B&:=&\frac{1}{2}(aB-J(B)a);\quad a\in \V;\quad B\in A\nn \\
a\dwedge B&:=&\frac{1}{2}(aB+J(B)a),
\ee
herewith we may decompose the Clifford product to
\be
aB&=&a\con B+a\dwedge B.
\ee
The contraction $\con$ is a graded derivative of degree -1, as
can be seen as follows (graded Leibnitz rule)
\be
a\con(bc)&:=&\frac{1}{2}(abc-J(bc)a)\nn \\
&=&\frac{1}{2}(abc-J(b)ac+J(b)ac-J(b)J(c)a)\nn \\
&=&(a\con b)c+J(b)(a\con c).
\ee
With $bc=1$ we have $a\con 1=2a\con 1$, so $a\con (R,C)=0$, from
which we could proof by induction the homogeneity of $a\con B_r$.
Obviously the contraction is linear, that is
\be
(\alpha X+\beta Y)\con A&=&\alpha X\con A+\beta y\con A.
\ee
These properties together state that $\con$ is an algebra
derivation. One can easy proof the useful formulas \cite{Loun}
\be
(u\wedge v)\con X&=&u\con(v\con (X))\nn \\
a\con (x_{i_1}\wedge \ldots \wedge x_{i_n})&=&\sum_{i=1}^n (-)^{i-1}
(a\con x_i)(x_1\wedge\ldots\wedge_{i-1}\wedge
x_{i+1}\wedge\ldots\wedge x_n)\nn \\
det(x_i\con x_j)&=&(x_n\wedge\ldots\wedge
x_1)\con(x_1\wedge\ldots x_n)\nn \\
&=&x_n\con(x_{n-1}\con\ldots(x_1\con(x_1\wedge\ldots\wedge x_n))\ldots).
\ee
Now the asymmetry of this result is obvious, and we may define
an arbitrary non degenerate bilinear form $B$ exactly as the
contraction. In a not necessarily orthonormalized
system of generating elements $e_i$ we have
\be
B&=&\brk{B_{ij}}=\brk{e_i\con e_j}.
\ee
The injection, introduced by Chevalley, $\bigwedge \V
\rightarrow CL(\V,Q)$, is of course known by physicists in the
disguise of the K\"ahler--Atiyah isomorphism.\footnote{
See \cite{Loun} for an account on that, and for a review on the
historical development.}

Therefore we have identified the Clifford algebra as a
subalgebra of $End(\oplus_n\bigwedge\V^{n})$, the endomorphism algebra of the
Grassmann algebra. A very explicit example will be given in
the appendix, in a manner closely related to the work of Lounseto.

Clearly, if we had chosen $J$ to be the common use involution
on $\V$, that is $J(\V)=-\V$, we would reobtain the original
formulas, with a symmetric bilinearform
\be
G_{ij}&=&\frac{1}{2}(e_ie_j-J(e_j)e_i)=\frac{1}{2}(e_ie_j+e_je_i).
\ee
Thus, if there exists a distinct involution of period two, we
have the desired extension.


We are now able to construct a new generating system of
the Clifford algebra, which is antisymmetric with respect to
the reversion, by using the corresponding wedge
product $\dwedge$.
\be
&\{e_0;e_{i_1};e_{i_1}\dwedge e_{i_2};e_{i_1}\dwedge
e_{i_2}\dwedge e_{i_3};\ldots\},\quad \forall i_n;\quad i_1<i_2<\ldots&
\ee
But now we have
\be
e_{i_1}\dwedge e_{i_2}&=&e_{i_1}e_{i_2}-e_{i_1}\con e_{i_2}
=e_{i_1}e_{i_2}-B_{i_1i_2},
\ee
which is not antisymmetric with respect to the reversion as one
can see as follows
\be
(e_{i_1}\dwedge e_{i_2})\til&=&(e_{i_1}e_{i_2}-B_{i_1i_2})\til\nn \\
&=&e_{i_2}e_{i_1}-B_{i_2i_1}+(B_{i_2i_1}-B_{i_1i_2})\nn \\
&=&e_{i_2}\dwedge e_{i_1}+(B_{i_1i_2}^T-B_{i_1i_2})\not=
e_{i_2}\dwedge e_{i_1}
\ee
Here $T$ means matrix transposition. To avoid such a situation,
and for establishing the reversion as the (hermitean) transpose
of the matrix representation, we are forced to choose the bi--
and multivectors in a definite way. With $i_1<i_2<\ldots$ we set
\be
e_{i_1i_2}&:=&\frac{1}{2}(e_{i_1}\dwedge e_{i_2}-e_{i_2}\dwedge
e_{i_1})\nn \\
&=&\frac{1}{2}(e_{i_1}e_{i_2}-B_{i_1i_2}-e_{i_2}e_{i_1}+
B_{i_2i_1})\nn \\
&=&\frac{1}{2}(e_{i_1}e_{i_2}-e_{i_2}e_{i_1})-\frac{1}{2}(
B_{i_1i_2}+B_{i_2i_1})\nn \\
&=&e_{i_1}\wedge e_{i_2}-F_{i_1i_2},
\ee
were $B$ is now splitted into symmetric and antisymmetric parts
$B=G_S+F_A$, with respect to the usual matrix transpose. We obtain
the following rules, utilizing now the reversion and the
{\it standard} involution.
\be
\frac{1}{2}(e_{i_1}e_{i_2}+(e_{i_1}e_{i_2})\til)&=&
\frac{1}{2}(e_{i_1}e_{i_2}+e_{i_2}e_{i_1})=G_{i_1i_2}\nn \\
\frac{1}{2}(e_{i_1}e_{i_2}-(e_{i_1}e_{i_2})\til)&=&
\frac{1}{2}(e_{i_1}e_{i_2}-e_{i_2}e_{i_1})=:e_{i_1i_2}\nn \\
&=&\frac{1}{2}(e_{i_1}\con e_{i_2}+e_{i_1}\wedge e_{i_2}
              -e_{i_2}\con e_{i_1}-e_{i_2}\wedge e_{i_1})\nn \\
&=&e_{i_1}\wedge e_{i_2}+F_{i_1i_2}\nn \\
e_{i_1i_2}\til&=&e_{i_2i_1}=-e_{i_1i_2}\nn \\
J(e_{i_1i_2})&=&e_{i_1i_2}.
\ee
A third order term will be given as
\be
e_{i_1i_2i_3}&=&\frac{1}{2}(e_{i_1}e_{i_2i_3}+e_{i_2i_3}e_{i_1})\nn
\\
&=&e_{i_1}\wedge e_{i_2}\wedge e_{i_3}
   +F_{i_1i_2}e_{i_3}+F_{i_2i_3}e_{i_1}-F_{i_1i_3}e_{i_2}.
\ee
If we would like to have the transposition to act trivial on the
matrix representation of the vector elements, we have to
introduce a dual set of generating elements.

We finish this section, by recalling the main consequences of the
analysis done, with respect to the application in the next section.

If there is a nonsymmetric part in the contraction, then the
usual multivectors are not the
desired algebraic elements. The nondiagonal part of the
contraction leads to a refined treatment of the algebraic
properties. The antisymmetric parts are incorporated in the
multivectorial structure, where as the symmetric part should be
handled with dual sets of generating elements\footnote{
The matrix transpose is only in this special situation
equivalent to the reversion.}.

By looking at this constructions, we are forced to introduce a new
kind of multivectors.
As a Clifford algebra, the two constructions prove to be
isomorphic, at least in the nondegenerate case. But the linear
subspaces $\oplus^n\bigwedge\V^{(n)}$ and $\oplus^n\mbox{span}\{
e_{i_1\ldots i_n}\}$ are quite different represented.

\section{Application to the nonlinear spinor field model}

Now we want to have a look at the vertex term of the nonlinear spinor
field theory. This should correctly be done in the functional
space formulation, which exhibits the structure quite more
clearly \cite{StBo}. For brevity and simplicity, we
will give our arguments direct on the level of the field operators.

The quantization of fermionic fields is in effect the
introduction of a Clifford algebra, or CAR algebra as in this
context usually named \cite{Itz,Brevier}.
\be
\{\Psi_K^\dagger, \Psi_{K^\prime}\}_+&=&\delta_{KK^\prime}
\ee
With our indexing $\Psi_I=\Psi_{K\Lambda}=\{\Psi_{K1}^\dagger;
\Psi_{K2}\}$ we have\footnote{
This is of course a special basis, we may call it a Witt
basis \cite{Brevier}. If the hermitean conjugation is the usual
one, then the connection to Fock space is very
close \cite{Vortrag}. Therefore we will expect to have a (in this
formulation) invisible antisymmetric part. So it is essential to
have non Fock--states.}
\be
\{\Psi_I,\Psi_{I^\prime}\}&=&\kla{\begin{array}{cc} 0&1\\1&0
\end{array}}_{\Lambda\Lambda^\prime}
\delta_{KK^\prime}
\ee
This relation can be rewritten in the form
\be
\Psi_I\con \Psi_{I^\prime}+\Psi_{I^\prime}\con \Psi_I&=&
2G_{II\prime}=\delta_{II^\prime}
\ee
which now can be extended to an arbitrary bilinear form $B$. We
obtain in this way the antisymmetric part, exhibiting a new term
\be
[\Psi_I,\Psi_{I^\prime}]&=&2F_{II^\prime}+2\Psi_I\wedge\Psi_{I^\prime}.
\ee
 From the Clifford algebraic point of view, this corresponds to
the scalar and bivector part, if we use the usual wedge product.

This entity is in fact related with the propagator of the theory.
\be
P_{II^\prime}&=&{}_{Ph}<0\vert T(\Psi_I\Psi_{I^\prime})\vert 0>_{Ph}\nn \\
&=&{}_{Ph}<0\vert \theta(t_I-t_{I^\prime})\Psi_I\Psi_{I^\prime}
                 -\theta(t_{I^\prime}-t_I)\Psi_{I^\prime}\Psi_I
                 \vert 0>_{Ph}.
\ee
For equal times we have
\be
P^t_{II^\prime}&=&{}_{Ph}<0\vert \Psi_I\Psi_{I^\prime}
                 -\Psi_{I^\prime}\Psi_I\vert
                 0>_{Ph,t=t^\prime}\nn \\
&=&{}_{Ph}<0\vert 2F_{II^\prime}+2\Psi_I\wedge \Psi_{I^\prime}
                \vert0>_{Ph,t=t^\prime}.
\ee
Now the $F_{II^\prime}$ are 'scalars', that is in field theory a
distribution valued function, and act {\it not} as operators.


Looking in this way at the vertex term, we have antisymmetric
products, and are free to choose the appropriate one, which
absorbs the additional terms, resulting in the normalordering
procedure. Of course, this should be done in such a way that
the transposition and reversion behave in the right way, but
here we will not bother about that\footnote{
See the remarks in the appendix.}.

By comparing \cite{StBo}
\be
{\bf :}\Psi_{I^\prime}\Psi_{I^\pprime}\Psi_{I^\ppprime}{\bf :}&=&
       \Psi_{I^\prime}\Psi_{I^\pprime}\Psi_{I^\ppprime}
       -P_{I^\prime I^\pprime}\Psi_{I^\ppprime}
       +P_{I^\prime I^\ppprime}\Psi_{I^\pprime}
       -P_{I^\pprime I^\ppprime}\Psi_{I^\prime}\nn \\
e_{i_1}\wedge e_{i_2}\wedge e_{i_3}&=&e_{i_1i_2i_3}
   -F_{i_1i_2}e_{i_3}+F_{i_1i_3}e_{i_2}-F_{i_2i_3}e_{i_1},
\ee
it is shown, that if we choose $P_{II^\prime}$ as
the antisymmetric part of the contraction, then we are forced to
introduce the normalordering terms in the field equation from
the beginning. This is, because we want the usual conjugation and
the multivectorial construction to hold in the algebraic and matrix case.

For the field equation this yields
\be
(\sum i\gamma^{\mu}\partial_{\mu}-m)_{II^\prime}\Psi_{I^\prime}&=&
gV_{I I^\prime I^{\pprime} I^{\ppprime}}
\Psi_{I^\prime}\wedge\Psi_{I^\pprime}\wedge\Psi_{I^\ppprime},
\ee
or
\be\label{stern}
&(\sum i\gamma^{\mu}\partial_{\mu}-m)_{II^\prime}\Psi_{I^\prime}
+gV_{I I^\prime I^\pprime I^\ppprime}\{
       P_{I^\prime I^\pprime}\Psi_{I^\ppprime}
      -P_{I^\prime I^\ppprime}\Psi_{I^\pprime}
      +P_{I^\pprime I^\ppprime}\Psi_{I^\prime}\}
=&\nn \\
&gV_{I I^\prime I^{\pprime} I^{\ppprime}}
\Psi_{I^\prime I^\pprime I^\ppprime}.&
\ee
Omitting now the interaction term (RHS of \ref{stern}) we are
left with a still singular equation, but now the singularity is
only the dynamical one. As proposed in the introduction, the
dynamical singularities may also be treated in an algebraic
manner, which will be shown elsewhere.

The Clifford algebraic point of view should of course be taken
from the very beginning.

\section{Conclusion}

With help of some results obtained by studying Clifford algebras
with non symmetric bilinear forms, we are able to understand the
process of normalordering in a new and deeper way. Without this
sort of tool, it seems hardly to be possible to recognize the
algebraic difference between $T$ and $N$ ordered transition
matrix elements. In fact they belong to quite different
multivector constructions. In ordinary treatments the vertex
normalordering is done ad hoc, simply motivated by obtaining an
afterwards finite theory.

At least in the computation of composites one has to expect the
appearance of nonsymmetric parts of the bilinear forms. This
stems from the antisymmetric constructions of the composite, in
which case the effective bilinearform should have such a part.

The next step is the observation, that the usually obtained
divergencies are related to the dynamical ones. Therefore it is
obvious, that they are irrelevant to the not yet well defined
theory, because they evaporate when the theory is regularized. A
posteriori the 'dot' procedure is thus justified. But the important
thing is, that we have, even in this case, to choose an other
timeordered equation. The construction itself gives the hint,
that we should start from the very beginning with Clifford
methods. Thereby Clifford analysis, or monogenetic
function theory, should us provide a finite theory, from first principles.

\section*{Acknowledgement}
The author acknowledge discussions with Dr. W. Pfister, and the
group of Prof. H. Stumpf, as Dr. Th. Borne for a critical
reading of the manuscript.

%
\section*{Appendix}
\begin{appendix}

In the Appendix an example is given, in the spirit of
Lounesto\cite{Loun}. Because all used quantities can only be
constructed explicitly in very low dimensional cases, we use
the Pauli algebra. It is well known and the smallest Clifford
algebra over the reals which exhibits a three--vector quantity.

The bilinear form is decomposed into symmetric and
antisymmetric parts, using matrix transposition. We have the
linear independent not normalized, not orthogonal set $\{e_1, e_2,
e_3\}$ spanning $\V$. The algebra is generated by
\be
&\{e_0;e_1,e_2,e_3;
       e_1\wedge e_2,e_2\wedge e_3,e_3 \wedge e_1;
       e_1\wedge e_2\wedge e_3\}.&
\ee
In this basis the bilinear form is
\be
B&=&\brk{B_{ij}}=\brk{e_i\con e_j}=\brk{g_{ij}}+\brk{f_{ij}}\nn \\
\brk{g_{ij}}^T&=&\brk{g_{ij}}\nn \\
\brk{f_{ij}}^T&=&-\brk{f_{ij}}.
\ee
Next we search for a matrix representation. This can be done
\cite{Loun} by Clifford multiplying from the right an algebra element
with all elements of the algebraic basis and expanding the result in
homogeneous parts. Those are written as columns of the matrices.
Matrix multiplication corresponds to the Clifford product. Of course
we have
\be
\brk{1}&=&\brk{\delta_{ij}},
\ee
and we calculate as an example $\brk{e_1}$
\be
e_1 1&=&e_1\nn \\
e_1 e_1&=& g_{11}\nn \\
e_1 e_2&=& e_1\con e_2+e_1\wedge e_2=g_{12}+f_{12}+e_1\wedge
         e_2\nn \\
e_1 e_3&=& g_{13}+f_{13}+e_1\wedge e_3\nn \\
e_1 (e_1\wedge e_2)&=&e_1(e_1 e_2 -e_1\con
e_2)=g_{11}e_2-(g_{12}+f_{12})e_1\nn \\
e_1 (e_2\wedge e_3)&=&(g_{12}+f_{12})e_3-(g_{13}+f_{13})e_2
                      +e_1\wedge e_2\wedge e_3\nn \\
e_1 (e_3\wedge e_1)&=&-g_{11}e_3+(g_{13}+f_{13})e_1\nn \\
e_1 (e_1\wedge e_2\wedge e_3)&=&g_{11}e_2\wedge e_3
                              +(g_{12}+f_{12})e_3 \wedge e_1
                              +(g_{13}+f_{13})e_1 \wedge e_2
\ee
The same can be done for the other elements, which yields
\be
\brk{e_1}&=&\brk{\begin{array}{cccccccc}
\x 0 &\x g_{11} &\x g_{12}+f_{12} &\x g_{13}+f_{13} &\x 0 &\x 0 &\x 0 &\x 0\\
\x 1 &\x 0 &\x 0 &\x 0 &\x -g_{12}-f_{12}&\x 0 &\x g_{13}+f_{13} &\x 0\\
\x 0 &\x 0 &\x 0 &\x 0 &\x g_{11} &\x -g_{13}-f_{13} &\x 0 &\x 0\\
\x 0 &\x 0 &\x 0 &\x 0 &\x 0 &\x g_{12}+f_{12} &\x -g_{11} &\x 0\\
\x 0 &\x 0 &\x 1 &\x 0 &\x 0 &\x 0 &\x 0 &\x g_{13}+f_{13}\\
\x 0 &\x 0 &\x 0 &\x 0 &\x 0 &\x 0 &\x 0 &\x g_{11}\\
\x 0 &\x 0 &\x 0 &\x-1 &\x 0 &\x 0 &\x 0 &\x g_{12}+f_{12}\\
\x 0 &\x 0 &\x 0 &\x 0 &\x 0 &\x 1 &\x 0 &\x 0 \end{array}}\nn
\ee
\be
\brk{e_2}&=&\brk{\begin{array}{cccccccc}
\x 0 &\x g_{21}+f_{21} &\x g_{22} &\x g_{23}+f_{23} &\x 0 &\x 0 &\x 0 &\x 0\\
\x 0 &\x 0 &\x 0 &\x 0 &\x-f_{22}&\x 0 &\x g_{23}+f_{23} &\x 0\\
\x 1 &\x 0 &\x 0 &\x 0 &\x g_{21}-f_{12} &\x -f_{23} &\x 0 &\x 0\\
\x 0 &\x 0 &\x 0 &\x 0 &\x 0 &\x g_{22} &\x -g_{12}+f_{12} &\x 0\\
\x 0 &\x -1&\x 0 &\x 0 &\x 0 &\x 0 &\x 0 &\x g_{23}+f_{23}\\
\x 0 &\x 0 &\x 0 &\x 1 &\x 0 &\x 0 &\x 0 &\x g_{12}-f_{12}\\
\x 0 &\x 0 &\x 0 &\x 0 &\x 0 &\x 0 &\x 0 &\x g_{22}\\
\x 0 &\x 0 &\x 0 &\x 0 &\x 0 &\x 0 &\x 1 &\x 0 \end{array}}\nn
\ee
\be
\brk{e_3}&=&\brk{\begin{array}{cccccccc}
\x 0 &\x g_{13}-f_{13} &\x g_{23}-f_{23} &\x g_{33} &\x 0 &\x 0 &\x 0 &\x 0\\
\x 0 &\x 0 &\x 0 &\x 0 &\x-g_{23}+f_{23}&\x 0 &\x g_{33} &\x 0\\
\x 0 &\x 0 &\x 0 &\x 0 &\x g_{13}-f_{13} &\x -g_{33} &\x 0 &\x 0\\
\x 1 &\x 0 &\x 0 &\x 0 &\x 0 &\x g_{23}-f_{23} &\x -g_{13}+f_{13} &\x 0\\
\x 0 &\x 0 &\x 0 &\x 0 &\x 0 &\x 0 &\x 0 &\x g_{33}\\
\x 0 &\x 0 &\x -1&\x 0 &\x 0 &\x 0 &\x 0 &\x g_{13}-f_{13}\\
\x 0 &\x 1 &\x 0 &\x 0 &\x 0 &\x 0 &\x 0 &\x g_{23}-f_{23}\\
\x 0 &\x 0 &\x 0 &\x 0 &\x 1 &\x 0 &\x 0 &\x 0 \end{array}}\nn
\ee
This 8 $\times$ 8 dimensional representation of the Pauli algebra
is not reducible to a real 4 $\times$ 4 or complex 2 $\times$ 2
one. The matrix transposition is not the reversion, because the
$[e_i]$ are not symmetric matrices. Also the trace is not an
algebraic invariant object, because there are elements with non
vanishing trace beside $[\delta^i_j]$, which means, that the
trace is not a projection on to the homogenouse part of degree zero.
So the matrix trace is not a linear form {\it in} the algebra,
because there are algebra elements except $\brk{\delta_{ij}}$ which are
not traceless. The trace is clearly a linear form on the matrix
representation, but into the field $(R,C)$ and not in the image
of the field in the algebra.

The volume element has nearly the bilinearform as entries in the
vector--vector block. The element $e_{123}$ reads
\be
\brk{e_{123}}&=&\brk{e_1e_2e_3-f_{12}e_3+f_{13}e_2-f_{23}e_1},
\ee
which yields a matrix not easy to display. The entries are
linear, quadratic and cubic functions of the $f_{ij}$ and
$g_{ij}$ parameters.

The same procedure can be done with the reordered basis,
belonging to the dotted wedge product or with respect to the basis
\be
&\{e_1,e_2,e_3;e_{123};
       e_0;e_{12},e_{23},e_{31}\}.&
\ee
ordered in odd and even elements. The vector elements read
\be
\brk{e_1}&=&\brk{\begin{array}{cccccccc}
\x 0 &\x 0 &\x 0 &\x 0 &\x 1 &\x -g_{12} &\x 0 &\x g_{31}\\
\x 0 &\x 0 &\x 0 &\x 0 &\x 0 &\x g_{11}  &\x -g_{13} &\x 0\\
\x 0 &\x 0 &\x 0 &\x 0 &\x 0 &\x 0 &\x g_{12} &\x -g_{11}\\
\x 0 &\x 0 &\x 0 &\x 0 &\x 0 &\x 0 &\x 1 &\x 0\\
\x g_{11} &\x g_{12} &\x g_{13}  &\x 0 &\x 0 &\x 0 &\x 0 &\x 0\\
\x 0 &\x 1 &\x 0 &\x g_{13} &\x 0 &\x 0 &\x 0 &\x 0\\
\x 0 &\x 0 &\x 0 &\x g_{11} &\x 0 &\x 0 &\x 0 &\x 0\\
\x 0 &\x 0 &\x -1&\x g_{12} &\x 0 &\x 0 &\x 0 &\x 0 \end{array}}\nn
\ee
\be
\brk{e_2}&=&\brk{\begin{array}{cccccccc}
\x 0 &\x 0 &\x 0 &\x 0 &\x 0 &\x -g_{22} &\x 0 &\x g_{23}\\
\x 0 &\x 0 &\x 0 &\x 0 &\x 1 &\x g_{12}  &\x -g_{23} &\x 0\\
\x 0 &\x 0 &\x 0 &\x 0 &\x 0 &\x 0 &\x g_{22} &\x -g_{12}\\
\x 0 &\x 0 &\x 0 &\x 0 &\x 0 &\x 0 &\x 0 &\x 1\\
\x g_{12} &\x g_{22} &\x g_{23} &\x 0 &\x 0 &\x 0 &\x 0 &\x 0\\
\x -1 &\x 0 &\x 0 &\x g_{23} &\x 0 &\x 0 &\x 0 &\x 0\\
\x 0 &\x 0 &\x 1 &\x g_{12} &\x 0 &\x 0 &\x 0 &\x 0\\
\x 0 &\x 0 &\x 0&\x g_{22} &\x 0 &\x 0 &\x 0 &\x 0 \end{array}}\nn
\ee
\be
\brk{e_3}&=&\brk{\begin{array}{cccccccc}
\x 0 &\x 0 &\x 0 &\x 0 &\x 0 &\x -g_{23} &\x 0 &\x g_{33}\\
\x 0 &\x 0 &\x 0 &\x 0 &\x 0 &\x g_{13}  &\x -g_{33} &\x 0\\
\x 0 &\x 0 &\x 0 &\x 0 &\x 1 &\x 0 &\x g_{23} &\x -g_{13}\\
\x 0 &\x 0 &\x 0 &\x 0 &\x 0 &\x 1 &\x 0 &\x 0\\
\x g_{13} &\x g_{23} &\x g_{33}  &\x 0 &\x 0 &\x 0 &\x 0 &\x 0\\
\x 0 &\x 0 &\x 0 &\x g_{33} &\x 0 &\x 0 &\x 0 &\x 0\\
\x 0 &\x -1&\x 0 &\x g_{13} &\x 0 &\x 0 &\x 0 &\x 0\\
\x 1 &\x 0 &\x 0&\x -g_{23} &\x 0 &\x 0 &\x 0 &\x 0 \end{array}}\nn
\ee
Which yields a much more convenient and symmetric form. If the
bilinear form is in the symmetric part diagonal, then this
representation becomes symmetric with respect to the trace. In
this case trace and reversion are identical. The
antisymmetric part has been absorbed fully in the multivectorial
construction.

A full satisfactory representation could be obtained by using a
dual set of generating algebra elements, to the above ones.
Therefore let the Volume element be
\be
E&:=&e_{123}=-E\til\nn \\
E^{-1}&=&frac{E\til}{E\til E}\nn \\
E\til E&=&det G=\vert G \vert.
\ee
Then we may construct
\be
e^i&:=&(-)^{i+1}e_{1\ldots i-1 i+1\ldots n} E^{-1}
\ee
which is a generalization to nonsymmetric bilinearforms of the
detailed results in \cite{Hest}.

Now the representation with the algebra basis $X^i$ yields via
$[e_i X^J]$ symmetric matrices, even if the symmetric part of
$B$ is nontrivial.

This form is the most distinguished and symmetric one.
Transposition and conjugation are the usual operations, but for
arbitrary $B$ the representation is still of dimension 8 $\times$ 8.

In this light, we have to change the 'quantization' process, to
use this dual elements. Therefore we should write
\be
\Psi^I\Psi_{I\prime}+\Psi_{I\prime}\Psi^I&=&\delta^i_{I\prime}.
\ee
But now the dual set depends on the possibly varying volume
element of the algebra, and makes the definition of 'creation'
and 'destruction' operators position dependent. We may hope to
get a better understanding of quantization on curved space in
this way.
\end{appendix}
%
%
%

%
%
\end{document}